\documentclass[conference]{IEEEtran}
\IEEEoverridecommandlockouts
\usepackage{cite}
\usepackage{amsmath,amssymb,amsfonts}
\usepackage{algorithmic}
\usepackage{graphicx}
\graphicspath{{fig/}}
\usepackage{textcomp}
\usepackage{xcolor}
\usepackage{url}
\usepackage{booktabs}
\usepackage{bm}
\usepackage{xspace}

\newcommand{\Z}{\ensuremath{\bm{z}}}          
\newcommand{\Za}{\ensuremath{\bm{z}_a}}       
\newcommand{\rvec}{\ensuremath{\bm{r}}}       
\newcommand{\xstate}{\ensuremath{\bm{x}}}     
\newcommand{\Hmat}{\ensuremath{\bm{H}}}       
\newcommand{\Rmat}{\ensuremath{\bm{R}}}       
\newcommand{\Jstat}{\ensuremath{J}\xspace}    
\newcommand{\dof}{\ensuremath{\nu}}           
\newcommand{\rtil}{\ensuremath{\tilde{\bm{r}}}}   
\newcommand{\evec}{\ensuremath{\bm{e}}}            
\newcommand{\wvec}{\ensuremath{\bm{w}}}            

\newtheorem{theorem}{Theorem}
\newtheorem{assumption}{Assumption}

\begin{document}

\title{Topology-Transferable Detection of Deep-Learning False Data Injection Attacks Through Residual-Energy Separation\\
}

\author{\IEEEauthorblockN{Yichen Liu, Hongyu Wu}
\IEEEauthorblockA{\textit{Dept. of Electrical and Computer Engineering} \\
\textit{Kansas State University}\\
Manhattan, KS, USA \\
\{ccwudi, hongyuwu\}@ksu.edu}
\and
\IEEEauthorblockN{Bo Liu}
\IEEEauthorblockA{\textit{School of Engineering and Applied Sciences} \\
\textit{Washington State University}\\
Richland, WA, USA \\
bo.liu1@wsu.edu}
}

\maketitle

\begin{abstract}
Deep learning (DL) false data injection (FDI) attacks can evade both bad data detection (BDD) and data-driven (DD) detectors in AC power system state estimation (SE), and DD detectors lose effectiveness after a topology change unless they are retrained. We argue that BDD fails in its decision rule, not in its statistic. A reconstruction-based DL generator pursues stealth by fitting the measurements closely, and because such a generator tends to act as a denoiser, the attacked residual energy can fall into the lower tail that BDD never inspects. The weighted-least-squares (WLS) residual applies the same projector construction to the noise and to the attack error, and under a spectral suppression condition we prove that the expected residual energies of normal and attacked measurements are separated by a topology-independent margin. Empirically, a rule learned on the residual energy at one topology transfers without retraining and keeps mean cross-topology $F_1$ above $0.98$ on the IEEE 14-bus, 39-bus, 57-bus, and 118-bus systems, while a raw-measurement DD detector collapses.
\end{abstract}


\section{Introduction}
\label{sec:intro}

\IEEEPARstart{S}{tate} estimation (SE) turns supervisory control and data acquisition (SCADA) measurements into the operating point used by contingency analysis and dispatch, which makes it a target for false data injection (FDI) attacks~\cite{zhang_smart_2021}. The standard screen, bad data detection (BDD), flags a snapshot when the energy of the weighted-least-squares (WLS) residual exceeds an upper $\chi^2$ threshold~\cite{schweppe_static_1970,monticelli_reliable_1983}. After Liu \emph{et al.}~\cite{liu_false_2011} showed that an attacker who knows the measurement Jacobian can leave this residual unchanged, the field turned away from the residual toward data-driven (DD) detectors that learn attacked patterns directly from the measurements~\cite{esmalifalak_detecting_2017,zhang_spatial-temporal_2022,liu_detecting_2025}.

Deep learning (DL) FDI attacks now cause two failures~\cite{nawaz_generative_2025}. First, they evade BDD: trained only on historical SCADA records, regression, adversarial, and autoencoder generators produce measurements that pass BDD and resemble normal operation~\cite{rahman_adversarial_2024,liu_gan-based_2022,zhao_controllable_2025,costilla-enriquez_attack_2023}. Second, DD detectors fail under reconfiguration: each line switch changes the map from state to measurements, so a DD detector trained at one topology reads a shifted distribution after the switch~\cite{dcvn2025,hu_training_2024}. Moving to the SE residual vector does not by itself help, because SE removes the state but not the topology. The operator holds no attack labels for a topology that has not yet occurred, and existing topology-adaptive DD detectors offer no guarantee on what survives the switch.

The two failures have complementary causes: BDD employs an appropriate test statistic but an ineffective decision rule, whereas DD detectors learn flexible decision rules from representations that are sensitive to topology changes. We therefore investigate the following question: what property of the SE residual distinguishes DL-FDI attacks from normal operation, and under what conditions does this distinction remain valid after reconfiguration? We show that the answer lies in the residual energy already computed by BDD, because the stealth objective of DL-FDI attacks leaves a distinct signature in this statistic. Although topology changes affect the residual energy, this signature persists across topologies and keeps normal and attacked measurements distinguishable. We therefore retain the BDD statistic and improve only its decision rule. The main contributions of this work are summarized as follows.

\begin{itemize}
\item \emph{Stealth by reconstruction makes the attack too clean.} A reconstruction-based generator pursues stealth against BDD by fitting the measurements closely. Because deep networks fit the reproducible structure of their data before its random component, a close fit tends to strip the noise as well. The stealth objective itself thus pushes the attacked residual energy below its normal level, into the lower tail that BDD never inspects.

\item \emph{Topology-uniform separation.} Normal measurement noise and the reconstruction error of the attack pass through the same kind of residual projector. We prove that if the attack error is less noisy than the measurement noise in every whitened direction, the expected residual energy of attacked measurements stays below the normal level by a margin set by the redundancy and the suppression level, not by the topology.

\item \emph{Residual-energy transferability across topologies.} We show that the scalar residual energy remains a transferable detection representation even when the measurement and residual-vector distributions change with the topology. On the IEEE 14-bus, 39-bus, 57-bus, and 118-bus systems, the defense trained at one topology keeps mean cross-topology $F_1$ above $0.98$ without retraining or new attack labels. On the 14-bus system it needs only ten labels per class, whereas raw-measurement and residual-vector DD detectors fail to transfer.
\end{itemize}

\section{State Estimation and DL-FDI Attacks}
\label{sec:background}

\subsection{WLS State Estimation and Bad Data Detection}
\label{subsec:acse}
At topology $T$, measurements $\Z\in\mathbb{R}^m$ satisfy $\Z=h_T(\xstate)+\evec$, where $\evec\sim\mathcal{N}(\bm{0},\Rmat)$, $\Rmat=\mathrm{diag}(\sigma_1^2,\ldots,\sigma_m^2)\succ0$, and $\xstate$ contains $n=2n_{\mathrm{bus}}-1$ voltage magnitudes and angles. The measurement map $h_T$ depends on topology, line parameters, and sensor placement; its Jacobian $\Hmat_T$ also depends on the operating point~\cite{schweppe_static_1970}. WLS minimizes $[\Z-h_T(\xstate)]^{\!\top}\Rmat^{-1}[\Z-h_T(\xstate)]$, leaving the residual $\rvec=\Z-h_T(\hat{\xstate})$ at the estimate $\hat{\xstate}$. In a fixed local linearization where $\Hmat_T$ has full column rank, whitening gives $\rtil=\Rmat^{-1/2}\rvec$ and $\wvec=\Rmat^{-1/2}\evec\sim\mathcal{N}(\bm{0},I_m)$, with
\begin{equation}
  \rtil=\bm{P}_T\wvec,\qquad \Jstat=\rvec^{\!\top}\Rmat^{-1}\rvec=\lVert\rtil\rVert^2=\wvec^{\!\top}\bm{P}_T\wvec,
  \label{eq:Jstat}
\end{equation}
where $\bm{P}_T=I_m-\tilde{\Hmat}_T(\tilde{\Hmat}_T^{\!\top}\tilde{\Hmat}_T)^{-1}\tilde{\Hmat}_T^{\!\top}$ and $\tilde{\Hmat}_T=\Rmat^{-1/2}\Hmat_T$. This orthogonal projector has rank $\dof=m-n$, the residual degrees of freedom. Under the linearized Gaussian model, $\Jstat\sim\chi^2(\dof)$~\cite{monticelli_reliable_1983}. Writing $\chi^2_p(\dof)$ for the $p$-quantile, BDD flags measurements with $\Jstat>\tau=\chi^2_{1-\alpha}(\dof)$ at false-alarm probability $\alpha$, leaving the lower tail untested because bad data inflate the residual.

\subsection{DL-FDI Attacks}
\label{subsec:fdia}
We consider a reconstruction-based DL attacker that trains a generator $G_\theta$ on historical measurements using
\begin{equation}
  \theta^\star = \arg\min_\theta\; \mathbb{E}\bigl\lVert G_\theta(\bm{u}) - \Z \bigr\rVert_2^2
  \label{eq:dl-fdia}
\end{equation}
and replaces $\Z$ with $\Za=G_{\theta^\star}(\bm{u})$ without using the physical measurement model. For a fully connected neural network (FCNN), $\bm{u}$ specifies an operating condition, such as bus voltage magnitudes~\cite{rahman_adversarial_2024,costilla-enriquez_attack_2023}; for an autoencoder (AE), it is the measurement vector~\cite{zhao_controllable_2025}. Relative to the state $\xstate_a$ represented by the attack, write
\begin{equation}
  \Za = h_T(\xstate_a) + \bm{\delta}_T, \qquad \bm{d}_T=\Rmat^{-1/2}\bm{\delta}_T.
  \label{eq:attack-model}
\end{equation}
To first order about $\xstate_a$, assuming a locally convergent WLS solution and a full-column-rank Jacobian, $\rtil_a=\bm{P}^a_T\bm{d}_T$ and $\Jstat^a=\bm{d}_T^{\!\top}\bm{P}^a_T\bm{d}_T$, where $\bm{P}^a_T$ is built as $\bm{P}_T$ but from $\Hmat_T(\xstate_a)$. It generally differs from the normal projector but has the same rank $\dof$. The same construction thus projects measurement noise under normal operation and reconstruction error under attack.

\section{Topology-Uniform Residual-Energy Separation}
\label{sec:theory}

\subsection{Stealth by Reconstruction Makes the Attack Too Clean}
\label{sec:clean}
A reconstruction-based attacker pursues stealth against BDD by making the attacked measurement satisfy the physical relations closely, which is what the objective~\eqref{eq:dl-fdia} rewards. Passing the upper threshold does not by itself require entering the lower tail, but nothing in the objective stops at the noise level, and the generator keeps fitting as far as its capacity allows.

What the generator can fit is the reproducible part of the data. For an FCNN whose input is independent of measurement noise, the expected squared error is minimized by the conditional mean $\mathbb{E}[\Z\mid\bm{u}]$. This equals the noiseless measurement when the input determines it. An AE receives noisy inputs, so its reconstruction objective alone does not ensure denoising. A bottleneck limits reconstruction capacity, and prior work on learning smooth structure and on denoising AEs suggests that reconstruction can suppress noise~\cite{arpit_closer_2017,rahaman_spectral_2019,vincent_extracting_2008}. This motivates the following sufficient condition, which we state as an assumption rather than a consequence of reconstruction training.

\begin{assumption}[Spectral suppression]\label{as:spectral}
Conditionally on each operating point, let $\bm{d}_T$ have mean $\bm{\mu}_T$ and covariance $\bm{\Sigma}^a_T$. Uniformly over the topologies and operating points considered, assume $\lambda_{\max}(\bm{\Sigma}^a_T)\le c<1$, $\lVert\bm{\mu}_T\rVert_2^2\le b$, and $b<(1-c)\dof$.
\end{assumption}

Whitened Gaussian normal noise, with zero mean and covariance $I_m$, is the boundary case $c=1$, $b=0$, for which~\eqref{eq:Jstat} gives the classical $\chi^2(\dof)$ law. The covariance bound limits variance in every whitened direction, and the bias bound ensures that a systematic reconstruction error does not eliminate the expected energy difference.

\subsection{Topology-Uniform Separation and the Learned Decision}
\label{sec:sep}
For fixed measurement count and redundancy, any two residual projectors satisfy $\bm{P}_{T'}=\bm{Q}\bm{P}_T\bm{Q}^{\!\top}$ for an orthogonal $\bm{Q}$. Thus, the normal residual covariance can change while the energy distribution remains $\chi^2(\dof)$. Attacked residuals need not be related by rotation. Their separation instead follows from Assumption~\ref{as:spectral}. All expectations below are conditional on a fixed operating point and its local projector; the uniform bounds also hold after averaging over operating points.

\begin{theorem}[Topology-uniform separation]\label{thm:sep}
Under Assumption~\ref{as:spectral}, for every topology $T$, with normal and attacked energies $\Jstat_T$ and $\Jstat^a_T$, and any fixed orthogonal projectors $\bm{P}_T$ and $\bm{P}^a_T$ of rank $\dof$,
\begin{equation}
  \mathbb{E}[\Jstat_T]-\mathbb{E}[\Jstat^a_T] \;\ge\; (1-c)\,\dof-b .
  \label{eq:margin}
\end{equation}
If $\bm{d}_T$ is also conditionally Gaussian, then $\Jstat^a_T$ is stochastically dominated by $\big(\sqrt{b}+\sqrt{c\,X}\big)^2$, where $X\sim\chi^2(\dof)$.
\end{theorem}

\begin{IEEEproof}
Normal operation gives $\mathbb{E}[\Jstat_T]=\operatorname{tr}\bm{P}_T=\dof$, and under attack $\mathbb{E}[\Jstat^a_T]=\operatorname{tr}(\bm{P}^a_T\bm{\Sigma}^a_T\bm{P}^a_T)+\bm{\mu}_T^{\!\top}\bm{P}^a_T\bm{\mu}_T\le c\,\dof+b$. For the tail, write $\bm{d}_T=\bm{\mu}_T+(\bm{\Sigma}^a_T)^{1/2}\bm{\xi}$ with $\bm{\xi}\sim\mathcal{N}(\bm{0},I_m)$. The matrix $(\bm{\Sigma}^a_T)^{1/2}\bm{P}^a_T(\bm{\Sigma}^a_T)^{1/2}$ has rank at most $\dof$ and eigenvalues at most $c$, so diagonalizing it gives $\lVert\bm{P}^a_T(\bm{\Sigma}^a_T)^{1/2}\bm{\xi}\rVert^2\le c\,X$ almost surely, with $X\sim\chi^2(\dof)$ formed from $\dof$ rotated coordinates of $\bm{\xi}$. With $\lVert\bm{P}^a_T\bm{\mu}_T\rVert\le\sqrt{b}$, the triangle inequality on $\lVert\bm{P}^a_T\bm{d}_T\rVert$ completes the proof.
\end{IEEEproof}

Both bounds depend on the projector rank rather than its orientation. Under the Gaussian condition, any rejection region containing a lower tail $\{\Jstat<\ell\}$ has detection probability at least $\Pr\big((\sqrt{b}+\sqrt{cX})^2<\ell\big)$ at every topology with the same $\dof$. A positive mean gap alone does not make this probability high; it reaches a target level $1-\beta$ when $\sqrt{b}+\sqrt{c\,\chi^2_{1-\beta}(\dof)}\le\sqrt{\ell}$, and with $\ell=\chi^2_{\alpha/2}(\dof)$ this requires small $c$ and $b$.

The defense therefore keeps BDD's statistic and learns only the decision, $\text{defense}(\Za)=g\big(\Jstat(\Za)\big)$, with SE run at the current topology and $g$ calibrated once. BDD is the one-sided member $g=\mathbb{1}\{\Jstat>\tau\}$, blind to the lower tail. A rule-based two-sided test needs no data and flags $\Jstat<\chi^2_{\alpha/2}(\dof)$ or $\Jstat>\chi^2_{1-\alpha/2}(\dof)$, an unsupervised rule replaces these quantiles by empirical ones from normal data, and a supervised rule trains a small multilayer perceptron (MLP) $g_\phi(\Jstat)$ by cross-entropy on labeled samples. For fixed $\dof$, the unchanged normal law preserves the false-alarm rate of any fixed rule, so each is reused at a new topology without new labels or retraining. The power bound applies to rules that reject the lower tail, and transfer of the learned MLP is evaluated empirically.

\section{Case Studies}
\label{sec:exp}

\subsection{Test Systems, Attacks, and Baselines}
\label{sec:exp-setup}
\textbf{Systems.} We use the IEEE 14-bus, 39-bus, 57-bus, and 118-bus systems, with $8000$ MATPOWER snapshots per topology generated from ERCOT load profiles and, for the 118-bus system, WECC load profiles~\cite{zimmerman_matpower_2020}. Each system has a reference topology and three reconfigurations by line removal ($\mathrm{rm}\,A\_B$), addition ($\mathrm{add}\,A\_B$), or both. Redundancy is fixed across topologies except in the 118-bus system. Its $\dof$ varies from $352$ to $360$, so the defense uses $\Jstat/\dof$ there, and this case tests transfer beyond the theorem's fixed-redundancy setting.

\textbf{Attacks.} The primary attack uses an FCNN to map voltage magnitudes to measurements~\cite{rahman_adversarial_2024}, while the secondary attack employs an AE with a latent dimension of $13$. Both are trained to minimize~\eqref{eq:dl-fdia}. Training a separate generator for each topology creates a challenging scenario for topology-dependent detectors.

\textbf{Baselines.} BDD uses $\alpha=0.05$. Five raw-measurement DD detectors are trained on classical FDI attacks and tested on DL-FDI attacks: a support vector machine (SVM) with a radial basis function (RBF) kernel, a random forest, logistic regression, $k$-nearest neighbors ($k$NN), and an MLP. For the representation comparison, identical MLPs are trained on DL-FDI attacks with either raw measurements (the raw-measurement DD detector) or residual energy (the supervised defense), using the same $200$ labels and $1200$ test samples per class. We report $F_1$, precision (Prec), and recall (Rec).

\subsection{Existing Detectors at a Fixed Topology}
\label{sec:exp-e1}
At the base 14-bus topology, four of the five DD detectors trained on classical FDI attacks achieve $F_1\ge0.989$ on classical attacks but $F_1=0$ on the DL-FDI attack (Table~\ref{tab:e1}). Logistic regression reaches only $0.092$ on the DL-FDI attack, and BDD detects none of it. Thus, strong performance on classical FDI attacks does not carry over to the evaluated DL-FDI attack, even without a topology change.

\begin{table}[!t]
\renewcommand{\arraystretch}{1.05}
\caption{Existing Detectors Against the DL-FDI Attack (IEEE 14-bus, Base Topology)}
\label{tab:e1}
\centering
\setlength{\tabcolsep}{4pt}
\begin{tabular}{lcccc}
\toprule
 & Classical & \multicolumn{3}{c}{DL-FDI attack} \\
\cmidrule(lr){3-5}
Method & $F_1$ & $F_1$ & Prec & Rec \\
\midrule
BDD            & --      & $0.000$ & $0.000$ & $0.000$ \\
SVM (RBF)      & $0.999$ & $0.000$ & $0.000$ & $0.000$ \\
Random forest  & $1.000$ & $0.000$ & $0.000$ & $0.000$ \\
Logistic reg.  & $0.563$ & $0.092$ & $0.142$ & $0.068$ \\
$k$NN ($k{=}5$) & $0.989$ & $0.000$ & $0.000$ & $0.000$ \\
MLP            & $1.000$ & $0.000$ & $0.000$ & $0.000$ \\
\bottomrule
\end{tabular}
\end{table}

\subsection{Cross-Topology Transfer}
\label{sec:exp-e2}
\begin{figure}[!t]
\centering
\includegraphics[width=0.88\columnwidth]{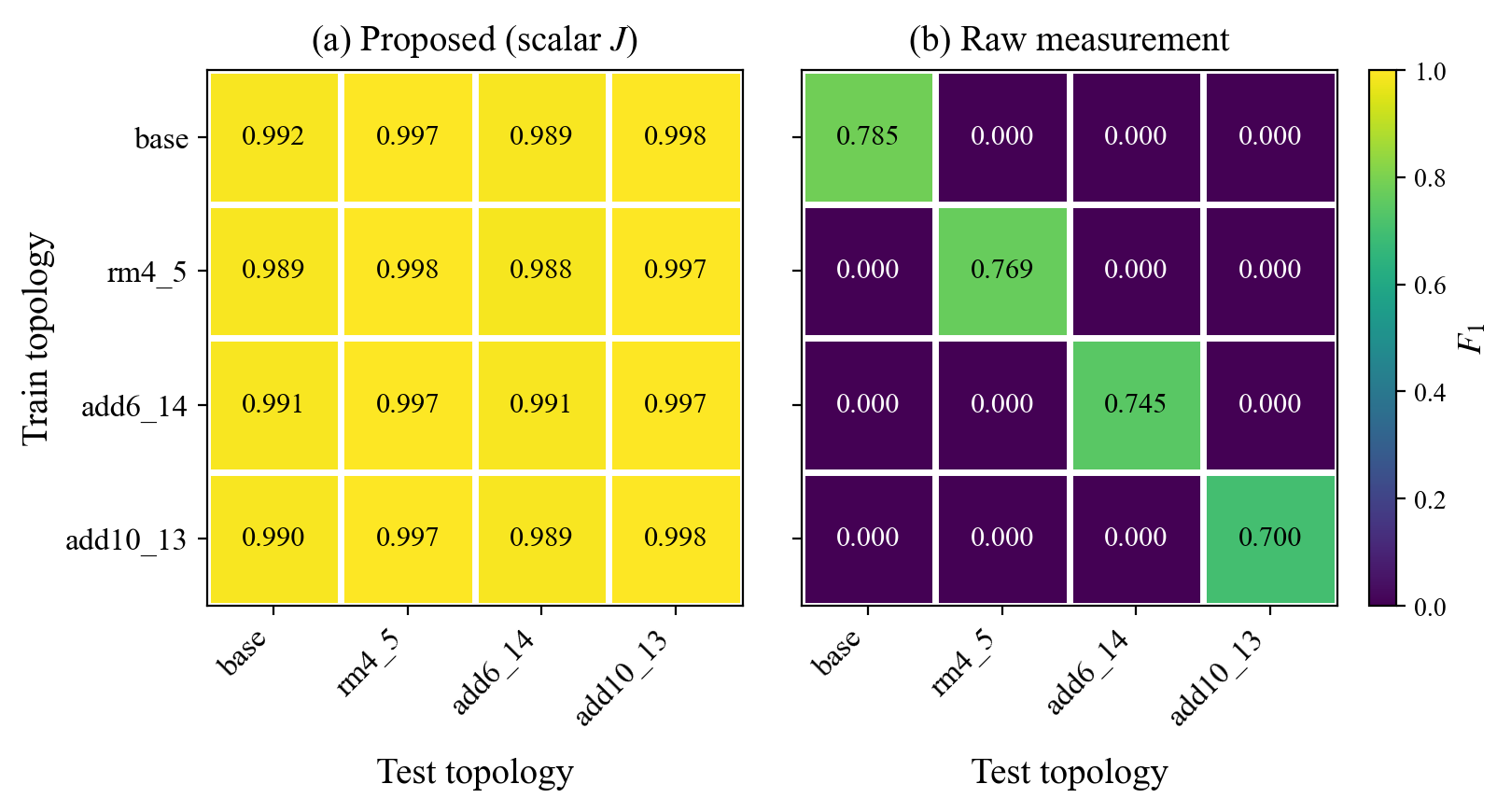}
\caption{Cross-topology transfer matrices ($F_1$) on the IEEE 14-bus system: (a) the supervised defense on $\Jstat$ and (b) the raw-measurement DD detector.}
\label{fig:e2heat}
\end{figure}

In the 14-bus transfer matrices, each row corresponds to training at one topology and testing at all four (Fig.~\ref{fig:e2heat}). The supervised defense achieves mean $F_1$ scores of $0.995$ at the training topology and $0.994$ across topologies. By contrast, the raw-measurement DD detector drops from $0.750$ to $0.000$ (Table~\ref{tab:e2}). A separate experiment with $200$ labels per class shows a similar limitation for the full residual vector: a residual-vector DD detector achieves $F_1=0.980$ at the training topology but only $0.002$--$0.441$ at the other three, whereas the defense remains above $0.994$. Without attack labels, the two-sided and empirical-quantile rules achieve $F_1$ near $0.97$, compared with $0.003$ for one-sided BDD.

\begin{table}[!t]
\renewcommand{\arraystretch}{1.05}
\caption{Mean Cross-Topology Transfer (IEEE 14-bus, FCNN Attack)}
\label{tab:e2}
\centering
\setlength{\tabcolsep}{3pt}
\begin{tabular}{lccc}
\toprule
Method & Same-topo.\ $F_1$ & Cross-topo.\ $F_1$ & Gap \\
\midrule
Defense, supervised    & $0.995$ & $0.994$ & $0.001$ \\
Defense, unsupervised  & $0.972$ & $0.972$ & $0.000$ \\
Defense, rule-based    & $0.971$ & $0.971$ & $0.000$ \\
BDD (one-sided)         & $0.003$ & $0.003$ & $0.000$ \\
Raw-measurement DD detector & $0.750$ & $0.000$ & $0.750$ \\
\bottomrule
\end{tabular}
\end{table}

\begin{figure}[!t]
\centering
\includegraphics[width=0.75\columnwidth]{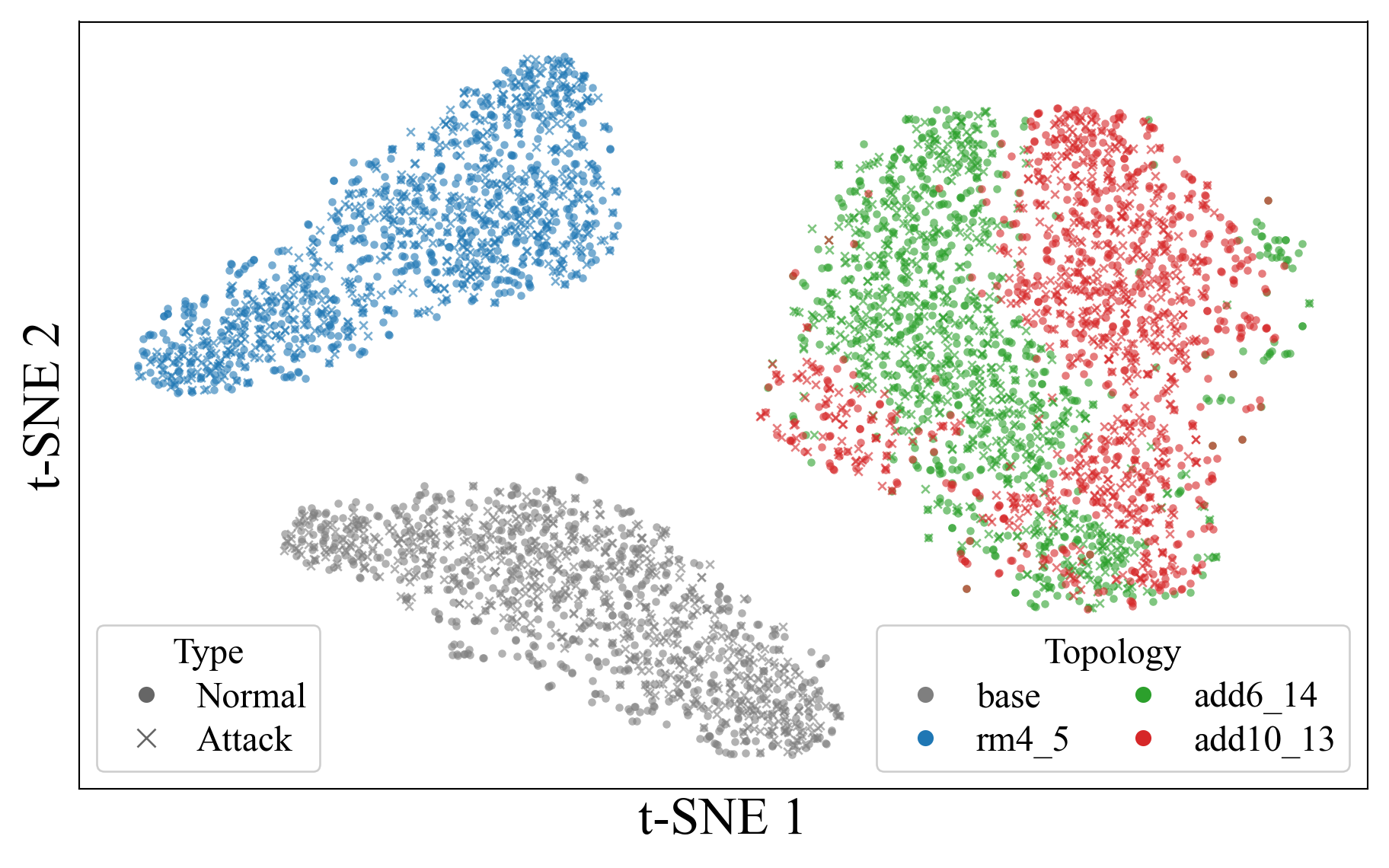}
\caption{t-SNE embedding of raw measurements at four topologies (IEEE 14-bus, FCNN attack).}
\label{fig:tsne}
\end{figure}

The t-SNE visualization in Fig.~\ref{fig:tsne} provides complementary evidence of topology dependence in raw measurements. Samples form four largely separated groups corresponding to the four topologies, while normal and attacked samples overlap within each group. This pattern is consistent with the raw-measurement DD detector's poor transfer performance in Fig.~\ref{fig:e2heat}.

\subsection{Residual Energy Across Topologies}
\label{sec:exp-jdist}
\begin{figure}[!t]
\centering
\includegraphics[width=0.8\columnwidth]{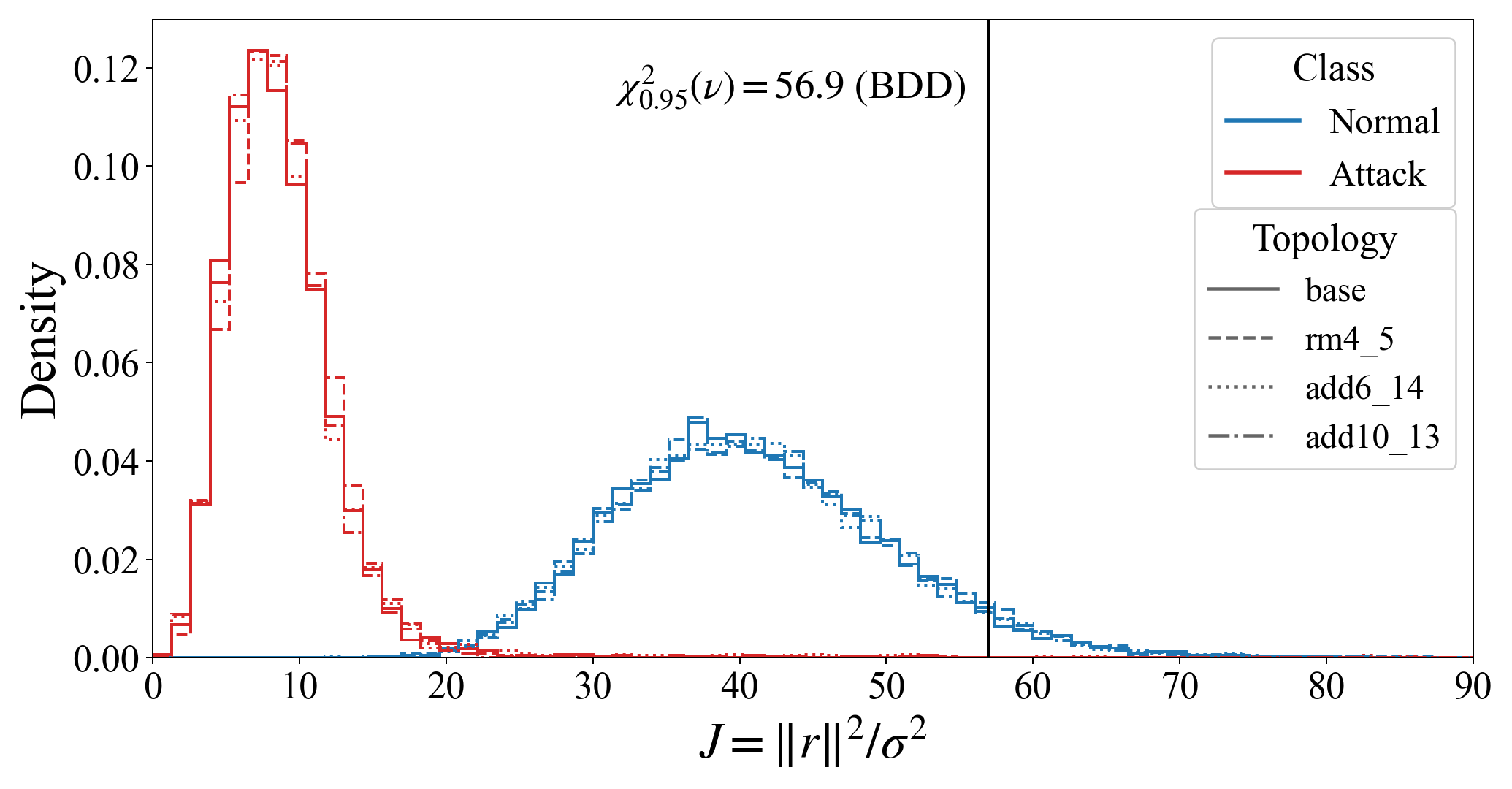}
\caption{Per-topology density of $\Jstat$ under normal operation and the FCNN attack (IEEE 14-bus).}
\label{fig:jdist}
\end{figure}

Figure~\ref{fig:jdist} shows that, across the four 14-bus topologies, normal residual energy concentrates near $\dof=41$, whereas attacked energy concentrates near $8$. This separation persists after reconfiguration, consistent with Theorem~\ref{thm:sep}, although the experiments do not estimate the theorem's spectral or bias bounds. The attacked-energy distributions also nearly overlap across topologies.

\begin{figure}[!t]
\centering
\includegraphics[width=0.8\columnwidth]{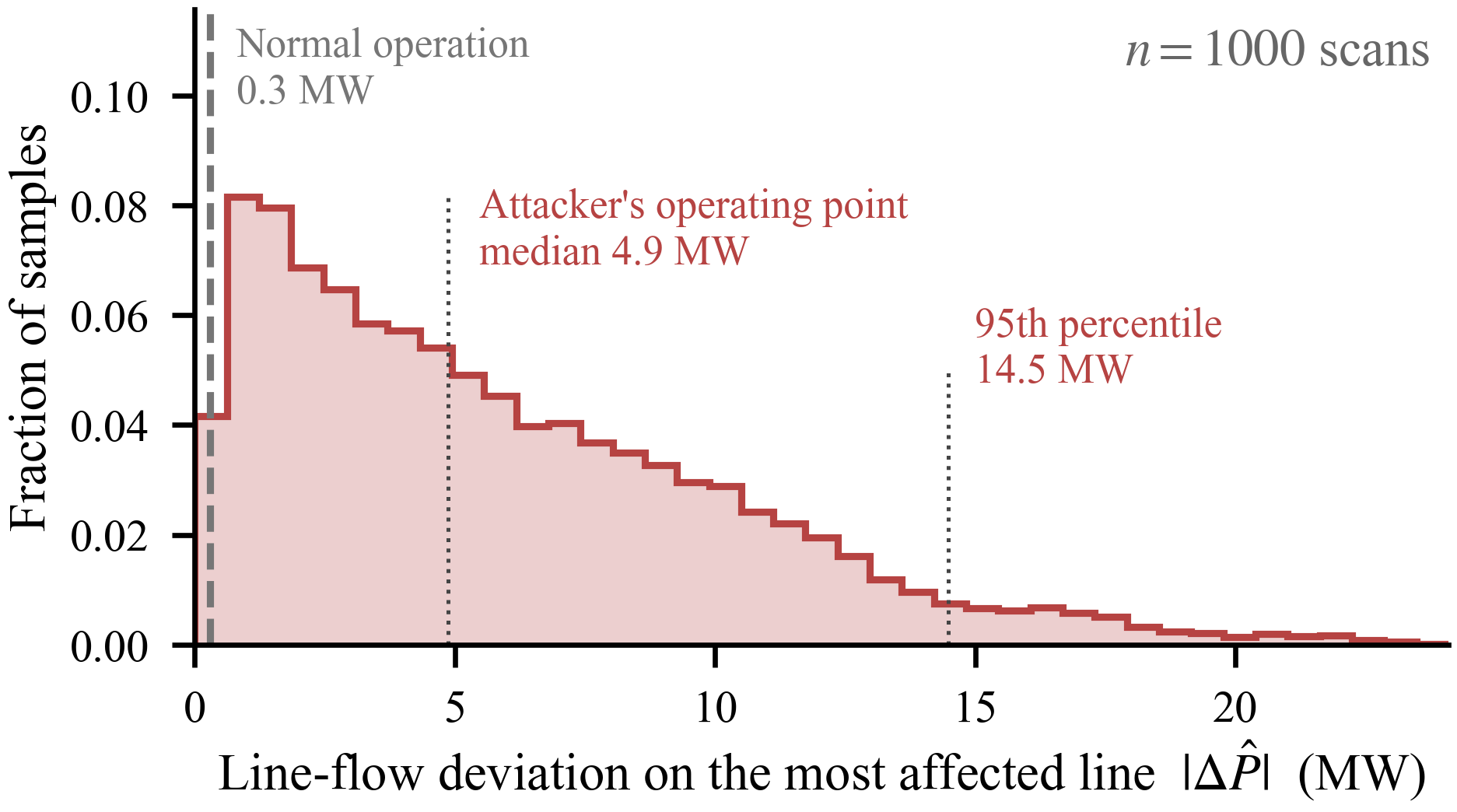}
\caption{Line-flow bias under the DL-FDI attack (IEEE 14-bus).}
\label{fig:impact}
\end{figure}

Low residual energy does not imply negligible attack impact. Figure~\ref{fig:impact} shows that, on the most affected line, the attack induces a median line-flow error of $4.9$~MW and a 95th-percentile error of $14.5$~MW, approximately $16\%$ and $50\%$ of the line flow, respectively. The corresponding normal-operation error is only $0.3$~MW.

\subsection{Training Topologies and Labels}
\label{sec:exp-e3}
\begin{figure}[!t]
\centering
\includegraphics[width=0.80\columnwidth]{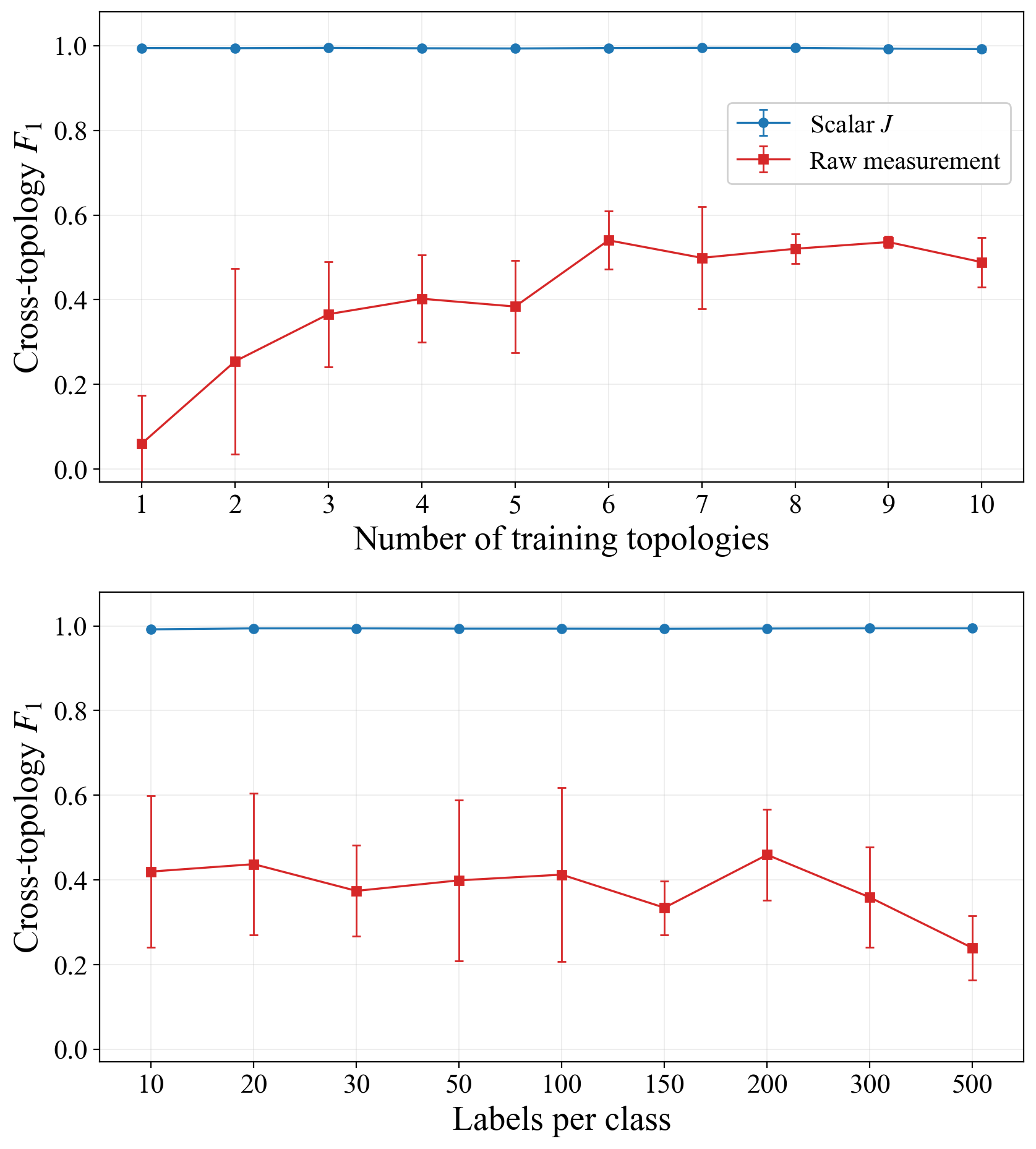}
\caption{Cross-topology $F_1$ versus the number of training topologies (top) and labels per class (bottom), IEEE 14-bus.}
\label{fig:ablation}
\end{figure}

Figure~\ref{fig:ablation} examines sensitivity to the number of training topologies and labeled samples. Each of five repetitions uses randomly selected training topologies and five held-out test topologies. With $200$ labels per class pooled across $K=1$--$10$ training topologies, the defense maintains $F_1$ near $0.99$, whereas the raw-measurement DD detector improves but plateaus near $0.5$. With a single training topology, the supervised defense reaches $F_1=0.992$ with only ten labels per class and remains near $0.994$ as the number increases to $500$. The raw-measurement DD detector shows no comparable gain.

\subsection{Generalization Across Systems and Attacks}
\label{sec:exp-e4}
To test whether the transfer generalizes beyond the 14-bus system, we apply the supervised defense to all four systems under the FCNN attack, and its mean cross-topology $F_1$ exceeds $0.98$ on each (Table~\ref{tab:e4}). The 118-bus mean is $0.986$, with individual transfer scores ranging from $0.961$ to $0.999$. To test generalization to a second attack, we replace the FCNN with the AE attack on three modified 14-bus topologies. The AE attack also yields low residual energy, and mean $F_1$ remains near $0.996$ both at the training topology and across topologies. Thus, the observed transfer is not limited to the FCNN attack.

\begin{table}[!t]
\renewcommand{\arraystretch}{1.05}
\caption{Mean Transfer Across Systems (FCNN Attack)}
\label{tab:e4}
\centering
\setlength{\tabcolsep}{4pt}
\begin{tabular}{lccc}
\toprule
System & BDD threshold & Same-topo.\ $F_1$ & Cross-topo.\ $F_1$ \\
\midrule
14-bus  & $56.9$  & $0.995$ & $0.994$ \\
39-bus  & $142.1$ & $0.997$ & $0.996$ \\
57-bus  & $201.4$ & $0.998$ & $0.999$ \\
118-bus & $401.0$ & $0.994$ & $0.986$ \\
\bottomrule
\end{tabular}
\end{table}

\subsection{Robustness to Measurement Noise}
\label{sec:exp-noise}
The $\chi^2(\dof)$ null distribution requires Gaussian noise, whereas the expected-energy gap in Theorem~\ref{thm:sep} depends only on the mean and covariance of the whitened noise and attack error. We test the 14-bus system with $\sigma=0.01$ under independent unit-variance Laplace, Student-$t(3)$, and uniform measurement perturbations, as well as spatially correlated Gaussian noise with correlation coefficient $\rho=0.5$. To avoid relying on a fixed $\chi^2(\dof)$ interval under these noise models, we use the unsupervised rule, estimating its quantile interval once at the base topology and applying it unchanged at the other three.

As shown in Table~\ref{tab:noise}, the three independent non-Gaussian noise models yield $F_1=0.954$--$0.979$ across the four topologies, close to the Gaussian reference and with little variation across topologies. Under correlated noise, $F_1$ falls to approximately $0.64$, consistent with the mismatch between the true noise covariance and the diagonal $\Rmat$ used for whitening. In a separate base-topology experiment, varying Gaussian noise magnitude from $0.5\%$ to $3\%$ keeps the normal-pass rate near $95\%$ and two-sided $F_1$ between $0.966$ and $0.979$.

\begin{table}[!t]
\renewcommand{\arraystretch}{1.05}
\caption{Cross-Topology $F_1$ Under Non-Gaussian Noise (IEEE 14-bus, Unsupervised Rule)}
\label{tab:noise}
\centering
\setlength{\tabcolsep}{5pt}
\begin{tabular}{lcccc}
\toprule
Noise & base & rm4\_5 & add6\_14 & add10\_13 \\
\midrule
Gaussian (ref.)  & $0.960$ & $0.967$ & $0.959$ & $0.964$ \\
Laplace          & $0.960$ & $0.963$ & $0.954$ & $0.966$ \\
Student-$t(3)$   & $0.968$ & $0.967$ & $0.959$ & $0.967$ \\
Uniform          & $0.971$ & $0.978$ & $0.966$ & $0.979$ \\
Correlated       & $0.639$ & $0.643$ & $0.645$ & $0.643$ \\
\bottomrule
\end{tabular}
\end{table}

\section{Conclusion}
\label{sec:conclusion}
This paper showed that BDD fails against DL-FDI attacks in its decision rule, not its statistic. A reconstruction attack that fits the measurements closely tends to remove the noise as well and can leave its residual energy in the lower tail. Because the normal residual energy keeps its $\chi^2(\dof)$ law at every topology with equal redundancy, a rule on this energy keeps its false-alarm calibration after reconfiguration, and under spectral suppression the expected gap between normal and attacked energy keeps a topology-independent lower bound.

On four IEEE test systems, such a rule trained at one topology kept mean cross-topology $F_1$ above $0.98$, while in the 14-bus comparisons, raw-measurement and residual-vector DD detectors failed to transfer. Future work includes noise-aware attackers, calibration under correlated measurement noise, and the conditions under which the attacked residual energy keeps an identical distribution across topologies.

\bibliographystyle{IEEEtran}
\bibliography{references}

\end{document}